\documentclass{eptcs}
\providecommand{\event}{ADG 2021} % Name of the event you are submitting

\usepackage{url}
\usepackage{cite}
\usepackage{graphicx}
\usepackage{framed}
\usepackage{breqn}
\usepackage{amsfonts}
\usepackage{listings}
\usepackage{color}
\usepackage{subfigure}
\usepackage{multicol}

\begin{document}

\def\orcidID#1{\unskip$^{\mbox{\href{https://orcid.org/#1}{\scriptsize{[#1]}} }}$}

\title{Online Generation of Proofs Without Words}

\author{Alexander Thaller 
\institute{Linz School of Education \\ Linz, Austria}
\email{alexander.thaller@geogebra.org}
\and
Zolt\'an Kov\'acs\orcidID{0000-0003-2512-5793}
\institute{The Private University College of Education of the Diocese of Linz \\
Linz, Austria} 
\email{zoltan@geogebra.org}
}

\def\titlerunning{Online Generation of Proofs Without Words}
\def\authorrunning{Thaller \& Kov\'acs}

\maketitle              % typeset the title of the contribution

\begin{abstract}
Understanding geometric relationships with little mathematical knowledge can
be challenging for today's students and teachers. A new toolset is introduced
that is able to create a proof without words by combining the benefits of the Geometric Deduction Database method
(to obtain a readable proof of a geometric statement) and the GeoGebra framework
(that makes it possible to export these data as an online applet in a simple way).
\end{abstract}

%\keywords{Geometry Deduction Database, Java Geometry Expert, GeoGebra, proof without words, %mathematics education}

\section{Introduction}
Automated proving in geometry is a celebrated research field that has been very successful over
recent decades, but mostly among researchers. World-wide use of applied methods,
mostly in schools, has just a limited success, even if remarkable applications
popularized the revolutional methods to obtain non-trivial proofs mechanically.
On one hand, the most successful methods provide only a yes/no answer if a proof is
sought (by occasionally extending the output with some non-degeneracy conditions).
On the other hand, even if some systems provide readable proofs, those computer programs
are not well-known, have a non-trivial user interface, or can be difficult to use
for a student who has no deeper knowledge in foreign languages.

Here we mention two important systems that have reached a high level of maturity. One of
them is the Java Geometry Expert (JGEX, \cite{Ye_2011}) which was developed around 2010. It uses the
Geometry Deduction Database (GDD, \cite{gdd}) method to obtain readable proofs. Unfortunately,
its source code was only opened in December 2016 and this long delay has prevented its
wider use and the option for feedback on further development. To date, fortunately, this project has
47 forks on GitHub and a couple improvements have already been contributed by external
developers. The Java Geometry Expert is available in a few languages,
however, its translation is still at an early stage.

The other system we focus on is GeoGebra \cite{gg3}. It has included support for automated reasoning
since 2015. It does not come with an option to show a readable proof: only
a yes/no-type answer is computed. On the other hand, its user interface is capable of creating
geometric constructions in a very simple way, even via a web interface.
Sharing online, copying and editing GeoGebra files are very common in the field of
education---in fact, GeoGebra became the de facto standard of constructing
geometric figures for the educational use. Its web portal \url{https:/www.geogebra.org}
provides millions of useful teaching materials as GeoGebra files in a freely
available way, ready for direct use immediately, in about 40 languages.
Clearly, 
free access to the GeoGebra system has resulted in a long-term advantage against its competitors.

In our contribution we show some on-going work how the outstanding proving
possibilities of JGEX can be integrated in GeoGebra's user interface to
help students and teachers to obtain readable proofs in a convenient way.
We focus on ease-of-use, but we also provide some technical details
how the two systems can be merged.

\section{The user workflow}
Assume that a user, say, a teacher creates a geometric construction by using GeoGebra's web version.
For example, a simple theorem is sketched that highlights some
property of planar Euclidean geometry, for instance, that the circumcenter,
the centroid and the orthocenter of a triangle are collinear.

By creating the circumcenter and the centroid, and drawing the line that joins them,
the teacher wants to check if the orthocenter is also an element of that line.
So the \textit{Relation} tool \cite{RelTool-ADG2014} will be selected, and then the line and the
orthocenter. First a numerical check is performed, and since it is positive,
a symbolic proof is computed. The current version of GeoGebra supports these described steps.
But right now we only get a yes or no answer if the theorem is true. We think that students
will benefit if GeoGebra is able to automatically generate a visual proof of the theorem.
We think that after displaying the yes/no type of answer of the symbolic proof, there should be another step,
where the user will be asked if a
proof without words is to be created. This proof will then be shown in a separate window of GeoGebra.

In fact, the created graphical proof is itself a GeoGebra file. It can be shared online,
copied and edited by the user. It can also be shown to students and other
teacher colleagues for presenting the idea behind the proof.

We highlight that the described process
does not require an expert user. Even if the teacher has no idea why
the theorem is true, the machine will still create a graphical, interactive applet
that fully describes the proof.

\section{Technical workflow}
In the background, GeoGebra's kernel is used to allow drawing the figure
and to verify the conjecture numerically. As a next step, GeoGebra's Automated Reasoning
Tools, a symbolic subsystem is invoked to quickly perform a symbolic proof.
Third, the GeoGebra construction will be automatically inserted into a
version of the JGEX system running in the background, that is able to load the GeoGebra file
and store the constructed geometric objects from the \texttt{.ggb} file as JGEX internals. After launching the GDD method,
the created data structure from JGEX that gives us the steps of the visual proof will be converted to a GeoGebra file format
to provide showing visual widgets (e.g.~sliders, animations, \LaTeX\ formulas)
in a modern look. 

Fortunately, from the technical perspective it is an achievable task to
merge JGEX and GeoGebra, because both systems are written in Java. Also,
GeoGebra's source code has been compilable into a web application
since 2011 (see \cite{GgM}). This makes it possible that the efforts of both developer groups
can be merged into a simple application. 

\section{Current state of our work}
Forks of JGEX and GeoGebra are maintained at the web pages
\url{https://github.com/halexus/Java-Geometry-Expert} and
\url{https://github.com/halexus/geogebra}. Currently JGEX is able to
load a GeoGebra file, and display the basic geometric objects and their constraints. You can now use GeoGebra as a construction tool and proceed with the proof inside of JGEX.
We also achieved to export the generated GDD-based proof as a GeoGebra file---in some specific cases. Those two
points were achieved by adding those functions to the JGEX software as two new dialogs, one to import and the other to export a \texttt{.ggb} file.
Both of the dialogs can be reached in the file menu of JGEX. 

The next step in this project will be the integration of this import and export functionality into GeoGebra and then automating the full process. In a last step
we will add the possibility to generate the proof with a graphical tool in GeoGebra's GUI.

We plan that the remaining steps will be implemented in a framework
of a master thesis by the first author and supervised by the second author. The implementation will then be freely available
to support learning and understanding of elementary geometry in a modern way.

%\section{Acknowledgements}

\bibliographystyle{eptcs}
\bibliography{kovzol,external}

\end{document}